\documentstyle[psfig]{mn}

\newif\ifAMStwofonts

\ifoldfss
  \ifCUPmtlplainloaded \else
    \NewTextAlphabet{textbfit} {cmbxti10} {}
    \NewTextAlphabet{textbfss} {cmssbx10} {}
    \NewMathAlphabet{mathbfit} {cmbxti10} {} 
    \NewMathAlphabet{mathbfss} {cmssbx10} {} 
  \fi
  \ifAMStwofonts
    \ifCUPmtlplainloaded \else
      \NewSymbolFont{upmath} {eurm10}
      \NewSymbolFont{AMSa} {msam10}
      \NewMathSymbol{\upi}     {0}{upmath}{19}
      \NewMathSymbol{\umu}     {0}{upmath}{16}
      \NewMathSymbol{\upartial}{0}{upmath}{40}
      \NewMathSymbol{\leqslant}{3}{AMSa}{36}
      \NewMathSymbol{\geqslant}{3}{AMSa}{3E}

    \fi
  \fi
\fi 

\ifnfssone
  \newmathalphabet{\mathit}
  \addtoversion{normal}{\mathit}{cmr}{m}{it}
  \addtoversion{bold}{\mathit}{cmr}{bx}{it}
  \newmathalphabet{\mathbfit} 
  \addtoversion{normal}{\mathbfit}{cmr}{bx}{it}
  \addtoversion{bold}{\mathbfit}{cmr}{bx}{it}
  \newmathalphabet{\mathbfss} 
  \addtoversion{normal}{\mathbfss}{cmss}{bx}{n}
  \addtoversion{bold}{\mathbfss}{cmss}{bx}{n}
  \ifAMStwofonts
    \ifCUPmtlplainloaded \else
      \UseAMStwoboldmath
      \makeatletter
      \new@mathgroup\upmath@group
      \define@mathgroup\mv@normal\upmath@group{eur}{m}{n}
      \define@mathgroup\mv@bold\upmath@group{eur}{b}{n}
      \edef\UPM{\hexnumber\upmath@group}
      \new@mathgroup\amsa@group
      \define@mathgroup\mv@normal\amsa@group{msa}{m}{n}
      \define@mathgroup\mv@bold\amsa@group{msa}{m}{n}
      \edef\AMSa{\hexnumber\amsa@group}
      \makeatother
      \mathchardef\upi="0\UPM19
      \mathchardef\umu="0\UPM16
      \mathchardef\upartial="0\UPM40
      \mathchardef\leqslant="3\AMSa36
      \mathchardef\geqslant="3\AMSa3E
    \fi
  \fi
\fi 

\ifnfsstwo
  \DeclareMathAlphabet{\mathbfit}{OT1}{cmr}{bx}{it}
  \SetMathAlphabet\mathbfit{bold}{OT1}{cmr}{bx}{it}
  \DeclareMathAlphabet{\mathbfss}{OT1}{cmss}{bx}{n}
  \SetMathAlphabet\mathbfss{bold}{OT1}{cmss}{bx}{n}
  \ifAMStwofonts
    \ifCUPmtlplainloaded \else
      \DeclareSymbolFont{UPM}{U}{eur}{m}{n}
      \SetSymbolFont{UPM}{bold}{U}{eur}{b}{n}
      \DeclareSymbolFont{AMSa}{U}{msa}{m}{n}
      \DeclareMathSymbol{\upi}{0}{UPM}{"19}
      \DeclareMathSymbol{\umu}{0}{UPM}{"16}
      \DeclareMathSymbol{\upartial}{0}{UPM}{"40}
      \DeclareMathSymbol{\leqslant}{3}{AMSa}{"36}
      \DeclareMathSymbol{\geqslant}{3}{AMSa}{"3E}
    \fi
  \fi
\fi 

\ifCUPmtlplainloaded \else
  \ifAMStwofonts \else 
    \def\upi{\pi}
    \def\umu{\mu}
    \def\upartial{\partial}
  \fi
\fi

\title{The Faint Sky Variability Survey II: Initial Results} 

\author[M. E. Everett et al.]
       {M. E. Everett$^1$, S. B. Howell$^1$, M. E. Huber$^1$,
        \newauthor
        P. J. Groot$^2$, P. M. Vreeswijk$^3$, J. van Paradijs$^{3,4}$
	\newauthor
	D. Davis$^5$, C. Neese$^5$, H. Scholl$^6$,
	\newauthor
	E. P. J. van den Heuvel$^3$, T. Augusteijn$^7$, H. B\"ohnhardt$^8$, 
	\newauthor
	P. A. Charles$^9$, T. J. Galama$^{10}$, E. Kuulkers$^{11,12}$, 
	\newauthor
	C. Kouveliotou$^{13}$, C. Moreno$^{14}$, G. Nelemans$^3$, 
	\newauthor
	R. Rebolo$^{15}$, R. G. M. Rutten$^7$, J. Storm$^{16}$, 
	\newauthor
	N. Tanvir$^{17}$, L. B. F. M. Waters$^3$, R. A. M. J. Wijers$^{18}$\\
        $^1$Astrophysics Group, Planetary Science Institute, 
            620 N. 6th Ave., Tucson, AZ, USA\\
	$^2$Harvard-Smithsonian Center for Astrophysics, 60 Garden St.,
            Cambridge, MA 02138, USA\\
        $^3$Astronomical Institute `Anton Pannekoek'/ CHEAF, 
            Kruislaan 403, 1098 SJ, Amsterdam, the Netherlands\\
	$^4$Physics Department, University of Alabama in Huntsville,
		Huntsville, AL, USA\\
        $^5$Planetary Science Institute, 620 N. 6th Ave.,
            Tucson, AZ, USA\\
	$^6$Observatoire de la C\^{o}te d'Azur, Nice, France\\
	$^7$Isaac Newton Group of Telescopes, Apartado de Correos 321, 
                Santa Cruz de La Palma, La Palma, Spain\\
	$^8$European Southern Observatory, Casilla 19001, Santiago 19, Chile\\
	$^9$Astronomy Department, University of Southampton, Southampton, UK\\
	$^{10}$Astronomy Department, California Institute of Technology, 
	        Pasadena, CA, USA\\
	$^{11}$Space Research Organization Netherlands,
		Sorbonnelaan 2, 3584 CA, Utrecht, the Netherlands\\
	$^{12}$Astronomical Institute, Utrect University, P.O. Box 80000,
	        3507 TA Utrecht, The Netherlands\\
	$^{13}$Marshall Space Flight Center (NASA), Huntsville, AL, USA\\
	$^{14}$Nordic Optical Telescope, La Palma, Spain\\
	$^{15}$Inst\'{\i}tuto de Astrof\'{\i}sica de Canarias, La Laguna, 
		Tenerife, Spain\\
	$^{16}$Astrophysikalische Institut Potsdam, An der Sternwarte 16,
	       14482 Potsdam, Germany\\
	$^{17}$Astronomy Department, University of Hertfordshire, 
 		Hertfordshire, UK\\
	$^{18}$Department of Physics and Astronomy, SUNY, Stony Brook, 
                NY 11794-3800, USA\\
       }
\date{}

\pagerange{\pageref{firstpage}--\pageref{lastpage}}
\pubyear{2000}

\begin{document}

\maketitle

\label{firstpage}

\begin{abstract}
We discuss the first results from the Faint Sky Variability Survey
(Groot et al. 2000).  The data consist of $V$-band light curves, $BVI$
colours, astrometry, and morphology information on several hundred
thousand point and extended sources in the magnitude range $V=17-25$.
We discuss the first 30 survey fields covering an area of 8.4 square
degrees towards moderate and high galactic latitudes.  We analyse the
quality of and discuss our differential photometry light curves.  We
employ statistical methods to select variable objects and present
example variable light curves.  The distribution of the colours and
magnitudes of point sources in the survey is discussed and compared to
galactic star count models.  Finally, we discuss our search for
trans-Neptunian objects in the FSVS fields observed towards ecliptic
opposition.
\end{abstract}


\section{Introduction}

The Faint Sky Variability Survey (hereafter FSVS) has been discussed
by Groot et al. (2000; hereafter paper I) and is part of the Wide
Field Survey.  The primary goals of the FSVS is to quantify the
photometric and astrometric variability of faint sources at moderate
to high galactic latitudes and to identify objects for detailed
follow-up observations.  The first two FSVS observing runs cover 8.4
square degrees to a 5-$\sigma$ limiting magnitude $V\sim25$ for point
sources, providing $V$-band light curves, $BVI$ colours, and
morphological and astrometric information for all objects in the
field.  The dataset of the FSVS allows us to detect rare stellar and
galactic populations exhibiting photometric variability, objects with
high proper motions or extreme colours, and trans-Neptunian objects.
In this paper we present initial results from the FSVS.  In Section~2
we review the observations that are discussed in paper I, in Section~3
we discuss the differential photometric techniques used in the FSVS,
and in Section~4 we show results.

\section[]{Observations}

We took the first observations for the FSVS on 1998 Nov. 16-21 UT, and
a second observing run was taken on 1999 May 11-17 UT.  The observing
strategy has been detailed in paper I so we describe it only briefly
here.  Each pointing of the telescope is centered on a different
``field'', consisting of the area imaged by the four Wide Field Camera
CCDs (each of the 4100x2048 pixel CCDs images an area of 0.072
sq. degrees for a total of 0.29 sq. degrees per field with a plate
scale of 0\farcs33~pixel$^{-1}$).  The total area of the sky covered
during these first two runs is 8.4 square degrees.  Each field is
observed 10-20 times with 10~minute exposures in the $V$-band
providing time sampling from $\sim13$ minutes up to, typically, 5
days.  Additionally, each field is observed once with exposures of 10
and 15 minutes respectively in the $B$- and $I$-bands to provide
colour information.  Astrometric plate solutions are found by taking
positions from the USNO-A2.0 Catalog (Monet et al. 1996).  Magnitudes
are put on a standard scale using observations of Landolt standard
star fields taken in succession with our data fields on the night that
we observe the field in $BVI$.

The 18 fields observed during the November run clustered in three
``areas'' of the sky, and the 12 observed during the May run were
clustered in two additional areas.  In this way, the FSVS samples a
variety of different galactic and ecliptic environments, each chosen
to avoid regions of obvious interstellar extinction so that distant
parts of the galactic halo could be observed.  Approximately one year
after each field is first observed it is scheduled for re-observation
with a single 10-minute V-band exposure to identify high proper motion
objects and those exhibiting long-term photometric variability.  The
areas observed during the first two FSVS observing runs are listed in
Table~1.

\begin{table*}
\begin{minipage}{115mm}
\caption{Areas observed in the FSVS up to May 1999}
\label{tab1}
\begin{tabular}{@{}lllll}
Fields & $\alpha_{2000}$ & $\delta_{2000}$ & $l$ & $b$ \\
1-6 & 23${\rm ^h}$ 44${\rm ^m}$ & +27\fdg2 & 105\degr & -33\degr \\
7-12 & 2${\rm ^h}$ 29${\rm ^m}$ & +14\fdg7 & 155\degr & -42\degr \\
13-18 & 7${\rm ^h}$ 54${\rm ^m}$ & +20\fdg7 & 200\degr & +23\degr \\
19-24 & 12${\rm ^h}$ 53${\rm ^m}$ & +27\fdg0 & 0\degr & +90\degr \\
27-30 & 16${\rm ^h}$ 25${\rm ^m}$ & +26\fdg5 & 45\degr & +43\degr \\
31,32 & 17${\rm ^h}$ 20${\rm ^m}$ & +26\fdg3 & 49\degr & +31\degr \\
\end{tabular}
\end{minipage}
\end{table*}

\section[]{Differential Photometry}

The light curves produced for every object in the FSVS are the result
of ensemble differential aperture photometry, wherein the observed
flux of each object is corrected for variations in atmospheric
transparency and seeing by comparing the observed fluxes to those of a
set of bright, non-variable calibration stars simultaneously observed
on the same CCD (hereafter referred to as reference stars).  The
techniques and application of such differential photometry are
discussed by Howell, Mitchell, \& Warnock (1988), Honeycutt (1992),
and Everett \& Howell (2000).  Because the FSVS dataset contains a
large number of sources, includes a wide range of object types (point
and extended sources), and was observed with a wide field and under
different observing conditions (seeing and transparency), it presents
a good test for our methods of photometry.  Note that here we are
concerned with the internal precision of each light curve, and this
depends only on comparisons between the reference stars and each
object.  This internal precision is distinct from our calibration of
the magnitude scale obtained through observations of separate
photometric standard star fields.

Our ability to perform precise differential photometric measurements
depends on reducing a combination of random and systematic
uncertainties to its lowest possible level.  Random errors arise from
the finite number of detected photons from the star and sky and
readout noise.  The magnitude of these random errors can be calculated
in a straightforward manner using Poisson statistics.  Systematic
errors arise mainly from imperfections in the instrumentation and
calibration which normally only become noticeable once the random
sources of noise are reduced to $<1$\ percent.  Sources of systematic
errors in the light curves include guiding errors, scattered light,
flatfielding uncertainties, and colour-dependent extinction.  It is
particularly important to understand the sources and behaviour of
these systematic errors because they can produce artifacts in light
curves that might otherwise be attributed to real variations.

The two-dimensional profiles of objects in the FSVS differ due to
intrinsic differences (galaxies vs. stars), extrinsic differences
(e.g., crowding), as well as instrumental or observational effects
(e.g., optical aberrations, seeing changes, guiding errors).  For
differential photometry we use the SExtractor package (Bertin \&
Arnouts 1996) with circular apertures that allow an exact comparison
of the fluxes from two objects having the same profile.  This assumes
that although the entire flux from each object is not contained within
the aperture, the same fraction of the flux will be.  This assumption
breaks down if one attempts to compare the fluxes from two objects
with different profiles.  For instance, these effects prevent us from
obtaining the high photometric precision light curves for galaxies
(which have unique profiles) that we have for stars.  Stars too may
have asymmetric profiles due to guiding errors or they may be
overlapped by the profiles of neighbouring objects.  In these cases we
either reject the data or interpret it with caution.

The systematic errors produced by differences and asymmetries in the
profiles of the reference stars and objects in the field may manifest
themselves as artificial ``variations'' in the objects' light curves.
Fortunately, we can distinguish these artifacts from true variations
because they are intercorrelated and often depend on variations in the
seeing (light curves of crowded or extended objects tend to exhibit
the same artificial patterns).  The following measures are taken to
reduce these artifacts and to reject problematic objects from the
dataset:  
\renewcommand{\labelenumi}{\arabic{enumi}.}
\begin{enumerate}
\item We reject the very few images taken with poor guiding.  
\item We measure magnitudes using four different circular aperture
sizes centered on each object.  The largest aperture is optimal for
the brightest stars (for which photometric precision is highest).  The
larger apertures reduce the adverse effects of slight differences
between the profiles of the reference stars and those of objects of
interest.  This is because a smaller fraction of the total flux falls
outside of large apertures, meaning any differences in the profile are
less important.  Other sources of uncertainty such as sky noise
dominate for fainter stars.  In this case, smaller apertures yield
better results.
\item We adopt a high threshold on the ``stellarity'' parameter to
distinguish point sources from extended sources (see paper I for more
discussion of the stellarity parameter).  This helps us to recognize
sources that are close to, but not quite, point-like (e.g., unresolved
binaries, compact galaxies).  Only light curves of uncrowded point
sources can be measured to high precision and realistically searched
for variability with errors calculated by Poisson statistics.
\item We choose the reference stars carefully by the method described
in paper I.  Since the noise in the (non-variable) reference star
light curves are uncorrelated, the variance of their mean light curve
is easily calculated and is significantly smaller than that of any
single object's light curve.  This ensures that no significant noise
is added to our light curves based on an uncertain reference star
correction.
\end{enumerate}

\section[]{Results}

After performing the basic data reductions as described in paper I, we
are in the position to analyse the data tables.  For each source, we
have a $V$-band light curve, information about the source profile in
each exposure (the stellarity parameter and a Gaussian fit to the
profile), and $B$- and $I$-band magnitudes.  In addition, coordinates
are found for each object.  In images where an object is not detected
by virtue of its faintness, we have taken our determination of the
plate limit as a lower limit to its magnitude.  Magnitude measurements
that could be in error (e.g., those of objects near bad pixels, the
edge of the CCD, crowded by neighbouring sources, or saturated) are
flagged and rejected before the analysis.

\subsection{Magnitudes and Colours}

A total of 500,000 objects was detected in the 30 fields (8.4
sq. degrees) observed during our first two FSVS runs.  The objects can
be classified according to their magnitudes, colours, variability, and
stellarity.  In Fig.~1 we show the magnitude distribution of point
sources and extended sources found in two different galactic regions
observed in the FSVS.  Here the point sources are taken to be those
with stellarity$>$0.8 and extended sources include those with
stellarity$<$0.2.  The thresholds for stellarity are somewhat
arbitrary, but for the bright sources these values distinguish
point-like from extended sources.  For fainter objects there is more
confusion distinguishing between point and extended sources due to
limited signal-to-noise and the diminishing angular size of galaxies
with distance.  This means that interpreting Fig.~1 at magnitudes
fainter than $V=22$ is difficult.  The histogram shows the magnitude
range over which we are sensitive ($V\sim17-25$), a range between the
brightest stars that remain unsaturated in the 10-minute exposures and
our faint detection limit.  We also see that the distribution of
extended sources rises more steeply at faint magnitudes reflecting the
tendency of galaxies to outnumber stars around $V=21$.

In panels (a) of Figs.~2 and 3 we show the magnitude distribution of
point sources at two different galactic latitudes alongside the
predicted distributions based on the Galaxy models of Bahcall \&
Soneira (1980).  The agreement between the observations and model is
generally good to as faint as $V=22$.  Towards the North Galactic
Pole, we observe an overabundance of point sources between $V=21$ and
$V=22.5$ relative to the model predictions.  Some of this excess can
be attributed to QSOs (Hartwick \& Schade 1990), but stars probably
contribute as well.  Also, in Figs.~2 and 3, the completeness limits
of the survey can be seen to be $V\sim22$ for our chosen stellarity
threshold of $>0.8$.  This is apparent because the observed starcounts
per magnitude turn over, whereas the model starcounts continue to
increase towards fainter magnitudes.  The detection limit is quite a
bit fainter than this completeness limit ($V\sim25$ compared to
$V\sim22$).  One reason is the fact that the stellarity is difficult
to measure for the faintest objects, and at low signal-to-noise, all
stellarity values tend towards 0.5.  Objects fainter than $V\sim22$
are in fact detected, but most have stellarity values less than 0.8.
Our completeness limit for simply detecting objects in these fields is
$V\sim24$.  We also note that the completeness and detection limits
depend on the observing conditions at the time the data were taken.
Some fields are observed under better conditions than others and,
naturally, exposures taken during the poorest observing conditions
limit our completeness.  The detections at $V\sim25$ are made on the
best exposures.  In a similar manner, the bright limit to the survey
depends on atmospheric transparency and seeing and this differs
between fields.  Therefore, the shape of the magnitude distribution at
the bright end ($V=15-17$) is statistically uncertain.  The panels (b)
in Figs.~2 and 3 show the Bahcall \& Soneira models broken down into
two stellar populations, namely a spheroid and disk component.  At the
North Galactic Pole, the FSVS contains primarily halo stars whereas at
mid-galactic latitudes (e.g., the $b=+23\degr\ l=200\degr$ fields
shown here), the disk stars dominate.

$BVI$ colours are measured from exposures of the same field taken
closely-spaced in time.  Thus, for large-amplitude variables, the
reported colours should indicate an approximate spectral type.  We
show a plot of $V-I$ vs. $B-V$ for point sources with stellarity$>$0.8
in Fig.~4.  The major grouping of points traces the main sequence, and
the majority of these sources are K and M dwarfs.  This can be seen by
the labels along the top of the plot that show the $V-I$ colors
expected for main sequence stars.  The colours for B-K dwarfs are from
Cox (2000) and those for (old disk) M dwarfs are from Leggett (1992).
Note that giants in the halo would have similar colours but most would
exceed the bright limit of the FSVS.  In Fig.~5 we show $V-I$
vs. $B-V$ for extended sources with stellarity$<$0.2.  Many galaxy
colours cluster around $V-I=1.5$ and $B-V=0.75$, falling below the
stellar main sequence on this plot.  Other galaxies have $BVI$ colours
that place them near the main-sequence or giant branch (the giant
branch and main-sequence are difficult to separate on this plot).
Note that Figs.~4 and 5 include only those sources detected in
every passband and thus, many of the objects with extreme colours are
not included since they are undetected at either $B$ or $I$.
Similarly, many of the faintest objects are not shown in either Fig.~4
or 5 because we chose not to include those with stellarity values
between 0.2 and 0.8.  Faint objects with such stellarity values are
difficult to classify.

\begin{figure}
\vspace{0pt}
 \caption{
Histograms showing the number of objects vs. magnitude in the 30 fields
of the FSVS.  The solid curve shows point sources (stellarity$>$0.8) and
the dotted line shows the same for extended sources (stellarity$<$0.2).
Both curves rise towards a completeness limit at $V\sim22$, with the
extended sources (mainly galaxies) dominating in numbers at magnitudes
fainter than $V\sim21$.  The faintness limit for actually detecting
objects is deeper than $V\sim22$; objects observed at low signal-to-noise
levels are difficult to classify by stellarity as the stellarity tends
towards 0.5.
}
\end{figure}

We show histograms of the distribution of objects in $B-V$ in Figs. 6
and 7.  We have included only point sources in these plots by
requiring stellarity$>$0.8 and $17<V<22$.  The magnitude limits are
chosen because the FSVS is complete within these limits for the point
sources as described above and as seen in Figs.~2 and 3.  The
distribution in $B-V$ has a bimodal distribution with the red peak
attributable to disk stars and the blue to halo stars.  Predictions
from the Bahcall \& Soneira models are plotted alongside the
observations and shown in terms of the disk and spheroid components.
Again, many features in the observed distributions are matched by the
models, but with less success for the mid-Galactic latitudes shown in
Fig.~6.  We did not attempt to fit the data with the model, of course,
and it is possible that the model parameters could be adjusted to
match the observations more closely.  For the mid-Galactic latitudes,
an overabundance of objects relative to the model predictions for
$0.4<B-V<1.2$ suggests that more halo stars are present than predicted
by the model.  The distribution in $B-V$ appears to be a better
discriminator between the two components than the distribution in $V$
shown in Fig.~2.

While most objects lie between $B-V = 0.4$ and 1.8, there are tails in
the distribution showing that a few objects have extreme colours.  The
model does not predict as many of the bluest objects as are observed.
The bluest objects will presumably include white dwarfs, CVs, and
emission-line objects, and are a prime target for our follow-up
classification spectroscopy.  Results of the spectra will be discussed
in future publications.

\begin{figure}
\vspace{0pt}
 \caption{
The number of point sources observed vs. $V$ towards a mid-galactic
latitude area at $b=+23\degr\ l=200\degr$ (fields 13-18) is shown
by the solid line in panel (a).  Here we require stellarity$>0.8$
to include only point sources.  The dashed line in panel (a)
shows the predicted $V$-distribution for the fields according to the
Galaxy model of Bahcall \& Soneira.  Panel (b) shows the total
number of stars predicted by the Bahcall \& Soneira model (solid
line), and the contributions to the total from the disk (dashed line)
and spheroid (dot-dashed line) components.
}
\end{figure}

\begin{figure}
\vspace{0pt}
 \caption{
The number of point sources observed vs. $V$ towards the North 
Galactic Pole (fields 19-24) is shown by the solid line
in panel (a).  A comparison with the Galaxy models of 
Bahcall \& Soneira is made.  See caption to Fig.~2 for details.
}
\end{figure}

\begin{figure}
\vspace{0pt}
 \caption{A plot of $B-V$ vs. $V-I$ for point sources (stellarity$>0.8$) 
found in the FSVS.  Most stars lie along the main sequence while a few 
outliers may be seen.  The $V-I$ colors of main sequence stars
are labelled along the top of the plot for reference.}
\end{figure}

\begin{figure}
\vspace{0pt}
 \caption{A plot of $B-V$ vs. $V-I$ for extended sources (stellarity$<0.2$) 
found in the FSVS.  Many of these galaxies cluster below the stellar
main sequence on this plot (see also Fig.~4) while others are coincident
with the colours of main sequence or giant stars.}
\end{figure}

\begin{figure}
\vspace{0pt}
 \caption{
Panel (a) shows the number of point sources observed vs. $B-V$ 
towards a mid-galactic latitude area at $b=+23\degr\ l=200\degr$ 
(fields 13-18) as a solid line.  Here we require stellarity$>0.8$
to include only point sources.  The dotted line shows the
predicted $B-V$ distribution for these fields according to the 
Galaxy model of Bahcall \& Soneira.  Panel (b) shows the same
model prediction as a solid line as well as the spheroid and
disk components as dot-dashed and dotted lines respectively.
}
\end{figure}

\begin{figure}
\vspace{0pt}
 \caption{
Same as for Fig.~6, but for fields 19-24 at the North Galactic Pole.
}
\end{figure}

\subsection{Photometric Variability}

The most unique aspect of the FSVS dataset is the variability
information available for each source.  The $V$-band light curves show
that photometric variability is detected in a few percent of the
sources.  These variations take place on all of the timescales
sampled, from $\sim$13 minutes to about five days.  The amplitudes of
variation also display a wide range, from about 5 millimagnitudes (or
the minimum amplitudes we could detect) to about one magnitude.

Our sensitivity to detecting photometric variations is a function of
the source brightness because sources detected with higher numbers of
counts can be measured (photometrically) to a higher precision.  To
demonstrate our photometric precision as a function of magnitude, we
plot the logarithm of the standard deviation measured in the data
points in each light curve versus the mean $V$-magnitude in Fig.~8.
Here we include only point sources by requiring a stellarity$>0.8$.
Since most sources do not vary, the standard deviations of their light
curves reflect the uncertainty in measuring each data point based on
counting statistics and these data trace out the locus of a precision
vs. magnitude function.  The data points in Fig.~8 are shown as either
dots or X's.  Light curves whose reduced $\chi^2$ statistic
($\chi^2_\nu$) for a constant fit to the light curve is greater than
10 are represented by the X's.  These stars are variable at a very
high level of confidence.  The dots represent stars with smaller
values of $\chi^2_\nu$.  Obviously on this plot variable sources tend
to lie above the non-variable sources.  There is no distinct dividing
line between the two, however, because variable observing conditions
mean some fields are observed to higher precision than others as well
as the fact that the calculation of $\chi^2_\nu$ is a weighted one
while that of $\sigma$ is unweighted.  The threshold of $\chi^2_\nu =
10$ is arbitrary, but adopting a high value prevents the
misidentification of non-variable objects as variable since sporadic
events (e.g., cosmic rays or asteroids in the aperture) can interfere
with our photometry.

To select a sample of variable objects from our data tables, we apply
statistical tests to each light curve.  Our first approach is to
calculate the $\chi^2_\nu$ statistic for a constant fit to each light
curve and to determine the likelihood that the observed variations are
not due to random fluctuations alone.  Sources that are found to be
variable at a high level of confidence are visually inspected on the
images for any unforeseen problems and their light curves are assigned
into ``bins'' by visual inspection and automated testing for different
types of variability.  We have targeted many of these variable objects
with follow-up classification spectroscopy and more detailed
photometry to determine their nature.

We show light curves for several variable point sources in Fig.~9.
Individual observations are spaced from approximately 13 minutes apart
(two consecutive 10-minute exposures with a 3-minute readout time) up
to 5 days.  Photometric variations are seen with both short- and
long-term behaviour, enabling us to create descriptive classifications
for each variable object (slow variations, rapid variations, low- or
high-amplitude variations, periodicity, etc.)

Periodic variations are actually difficult to detect given the limited
number of datapoints in the FSVS light curves.  However, visual
inspection of the light curves does show a few suggesting periodic
variations.  Several types of periodic and semi-periodic variable
stars are likely to show up in our dataset.  These include
chromospherically active late-type dwarfs, which show modulations in
time with their rotational period as spots move into and out of the
observer's direction, eclipsing binaries of many types, pulsating
variables, and interacting binaries.

RR Lyrae stars are likely to be present in the FSVS and good
candidates are drawn from those stars having variability on hour time
scales and with colours consistent with this class (panel (b) in
Fig.~9 shows the light curve of a candidate RR Lyr star).  Follow-up
time-series photometry can confirm their type and, from their
magnitudes, their distances can be estimated.

Since the FSVS is very rich in late-type dwarfs, it is therefore also
likely to contain a large number of spotted variables.  Indeed we see
many stars having slow variations over a 5-day time interval with
amplitudes ranging from several tenths of a magnitude down to the
smallest amplitudes we can detect.  Panels (a) and (d) in Fig.~9 show
light curves with this type of behaviour.  An analysis of variability
versus colour (and by implication, spectral type) will show how often
this affects the different late-type dwarf populations.

Without the variability information in the FSVS, rare classes of
objects having colours comparable to those of ``normal'' stars or
other objects may go unidentified.  For example, one of the main goals
of the FSVS is to search for the existence of faint, low mass-transfer
rate cataclysmic variable stars (CVs are interacting red dwarf-white
dwarf pairs; see Warner 1995).  CV candidates would be selected on the
basis of rapid variability.  The cataclysmic variable GO Comae was
observed in field 23 and its light curve is shown in panel (e) of
Fig.~9.  The colours are relatively blue ($B-V=-0.2$ and $V-I=0.9$),
but this object is most effectively identified via its rapid and
high-amplitude variations.  Much of the variation in this object is
attributable to flickering, although a possible 95-minute period may
produce some variations as well (Howell et al. 1990).  Candidate CVs
will be targeted for follow-up spectroscopy and confirmation would be
made by detecting the characteristic broad emission lines in their
spectra.

\begin{figure}
 \vspace{0pt}
 \caption{
The logarithm of the observed standard deviation in the light curves
of FSVS point sources (stellarity$>0.8$) vs. their mean $V$-magnitudes
for all 30 fields listed in Table~1.  Only light curves with 8 or more
error-free detections are included.  Light curves with $\chi^2_\nu>10$
are high probability variables and are shown as X's.  The standard
deviations found for the remaining sources are shown as dots.
}
\end{figure}

\begin{figure}
 \vspace{0pt}
 \caption{
Light curves of variable point sources.  For panels (a) through (d),
the spacing between consecutive data points is $\sim43$ minutes on the
first night and 13 minutes on the last night.  The CV star in panel
(e) (GO Com) is observed in the same manner as the others on the first
night and with exposures 13 minutes apart on the second night.  For
panels (a) through (d) the zeropoint in time is set to HJD=2451134 and
for panel (e) the zeropoint in time is set to HJD=2451313.  Error bars
in magnitude are shown in panels (a) and (d).  In panels (b), (c), and
(e), the errorbars are the same size or smaller than the points.
}
\end{figure}

\subsection{Trans-Neptunian Objects}

We searched three ecliptic opposition fields taken during the November
1998 run for trans-Neptunian objects (TNOs, also known as Kuiper Belt
objects), using a moving-object detection code written by one of us
(HS).  To detect TNOs, fields near opposition are needed in order to
distinguish TNOs from main-belt asteroids and other objects based on
their motion.  We found no TNO candidates in these fields.

We expected to find one known TNO, 1997~CS29, within the search area.
The predicted position for this TNO fell within one of our observed
ecliptic opposition fields on 1998 Nov 16 UT.  However, 1997~CS29 was
not found, probably due to a bright saturated star located at its
expected position.

Although the limiting magnitudes of many of the images from the
November run were as faint as $V = 25$, for detection of moving
objects, three, or preferably four images at different times need to
be intercompared.  The variable observing conditions during the
November run resulted in the sets of three or four exposures having
varying limiting magnitudes, and our detection sensitivity is limited
by the quality of the poorest image in the set.  As a result, the
effective limiting magnitude for our search was $V = 23.5 - 24$.

Given this limiting magnitude, over the 0.81 square degrees searched,
we would have expected to find 0.5 - 1 TNOs on average, based on the
TNO luminosity function determined by Gladman et al. (1998).  We used
the mean $V-R$ of known TNOs to apply the $R$ luminosity function of
Gladman et al. to our $V$ observations.  This is consistent with our
negative result.  In better weather conditions we can expect
consistent limiting magnitudes of $V > 25.5$ (similar to the best of
our November 1998 frames), and the expected number of TNOs then rises
to several per field.  Thus we will continue to search future FSVS
opposition fields for TNOs.

\section{Conclusions}

The Faint Sky Variability Survey is an ongoing deep, wide-field,
multi-colour, time-series CCD survey towards moderate and high
galactic latitudes.  Hundreds of thousands of objects in the magnitude
range $V=17-25$ are observed for classification on the basis of $BVI$
colours, temporal and astrometric variability, and morphology.
Analysis of the FSVS light curves reveals variable objects of many
types, objects with unusual colours, and will soon be extended to
include high proper motion stars.  Interesting populations have been
observed spectroscopically and will be discussed in future
publications.

\section*{Acknowledgments}

PJG and PMV are partially supported by NWO Spinoza grant 08-0 to Ed
van den Heuvel.  PJG is currently a CfA fellow.  SBH acknowledges
partial support of this project by NSF grant AST9818770.  MEH is
partially supported by the Wyoming Space Grant Consortium NASA grant
\#NGT-40008.  The FSVS is based on observations made through the Isaac
Newton Groups' Wide Field Survey Programme with the Isaac
Newton Telescope operated on the island of La Palma by the Isaac
Newton Group in the Spanish Observatorio del Roque de los Muchachos of
the Instituto de Astrofisica de Canarias.

\bsp

\label{lastpage}

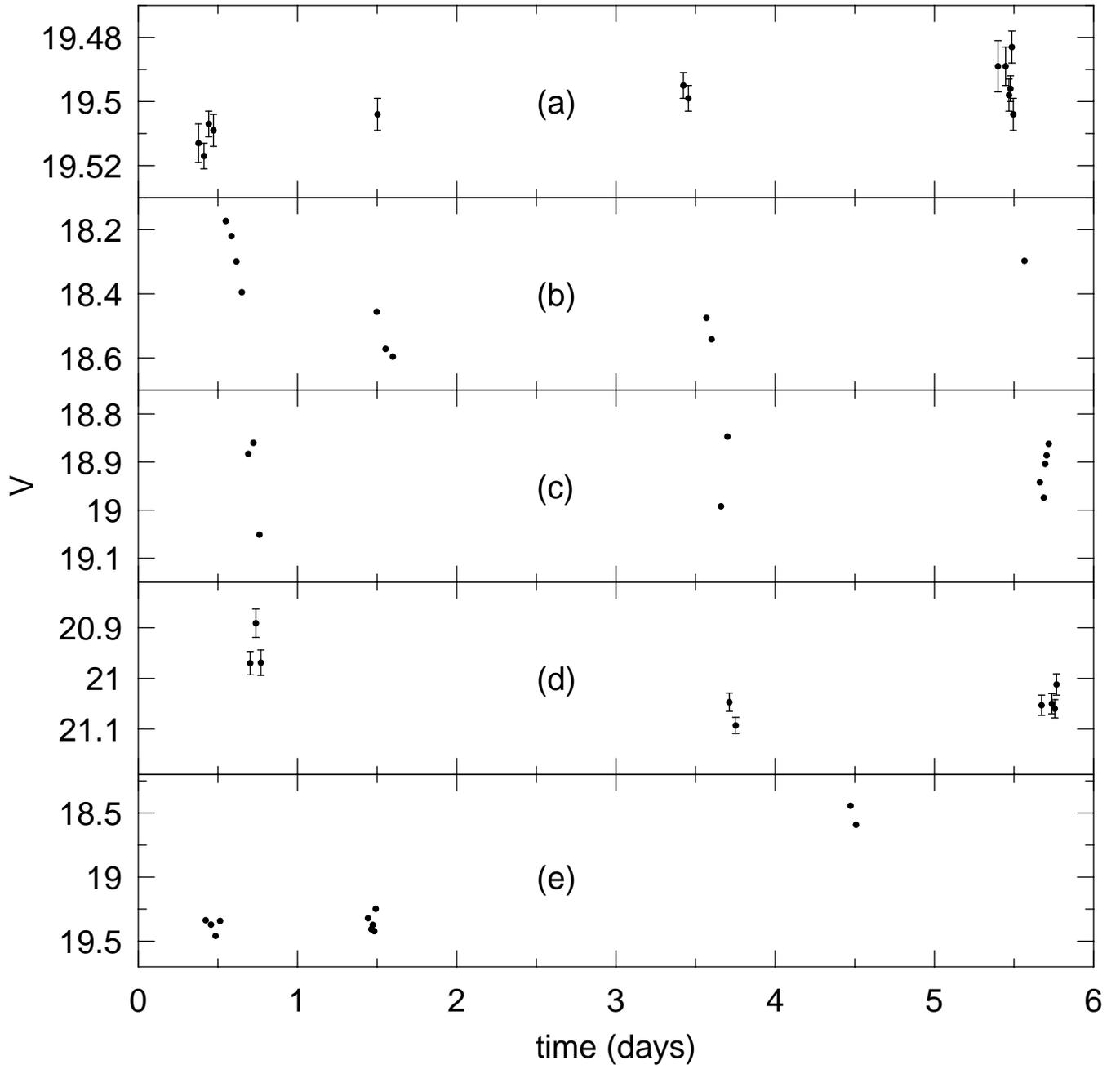
\end{document}